# Tackling Diversity and Heterogeneity by Vertical Memory Management*

Lei Liu (SKL, ICT)

## Abstract

*Existing memory management mechanisms used in commodity computing machines typically adopt hardware based address interleaving and OS directed random memory allocation to service generic application requests. These conventional memory management mechanisms are challenged by contention at multiple memory levels, a daunting variety of workload behaviors, and an increasingly complicated memory hierarchy. Our ISCA-41 paper proposes vertical partitioning to eliminate shared resource contention at multiple levels in the memory hierarchy. Combined with horizontal memory management policies, our framework supports a flexible policy space for tackling diverse application needs in production environment and is suitable for future heterogeneous memory systems.*

## 1. Introduction

Efficiently utilizing shared resources in the memory hierarchy such as Last Level Cache (LLC) and main memory is at the core of constructing high performance multi-core machines. To date, the most common mechanism of memory and cache sharing used in memory controllers in commodity machines is based on generic address interleaving, where the physical address of a memory request determines which LLC set and DRAM bank the request is serviced. Previous studies [1,3,7,8] indicate that this simple approach can cause significant memory/cache interference as multiple threads share the same DRAM banks and cache sets. Additionally, this approach is entirely oblivious to application and architecture characteristics, thus fail to efficiently use memory resource on modern computing systems with increasing diversity and heterogeneity.

### 1.1 Challenges for Existing Memory Management

Existing memory management approaches [6,8,9,11] typically focus on horizontally optimizing a single level in the memory hierarchy and have the following drawbacks:

**(1) Contention at different memory hierarchy:** Shared resources (i.e., LLC and DRAM) by multiple threads lead to contention at multiple levels in the memory hierarchy. Past efforts [7,8,9] focus on horizontally partitioning and managing LLC or DRAM banks to minimize contention at a single level. However, the contention problem has never been addressed for all levels in the memory hierarchy (except per-core private caches that do not suffer from inter-thread interference) at the same time. To completely eliminate interference in the memory hierarchy, a new approach is needed to vertically combine contention elimination techniques on multiple levels.

**(2) Single policy management:** Existing memory management in operating system (OS) is largely single policy based, which fails to support flexible and effective memory allocation with respect to different applications' sharing and capacity needs. As disparate non-volatile memory technologies are emerging and evolving to more sophisticated and hybrid memory systems [10,17], adaptive and reconfigurable policies are needed to manage the heterogeneity in terms of retention, access speed, fault tolerance, and energy efficiency. In such heterogeneous memory environments, single policy management can result in significant resource underutilization.

**(3) Application obliviousness:** Emerging workloads contain

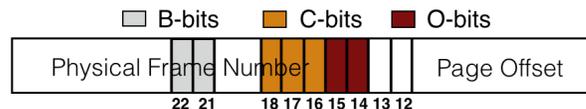

Figure 1. Address mapping from the view of OS and three categories of color bits on a typical multicore machine.

numerous applications and exhibit diverse and dynamic behaviors. Our results demonstrate that servicing all applications using one simple generic policy in a program-oblivious way often results in inter-program perturbation, resources thrashing, poor memory/cache utilization, and, consequently, degraded performance. Therefore, an intelligent system that can "understand" application behaviors is extremely important, especially in complicated computing environments such as cloud and data center, where a large number of applications are running and sharing memory resources [1,6,16,18].

### 1.2 Going Vertical in Memory Management

Our ISCA-2014 paper proposes a novel solution to meet the aforementioned challenges. (1) To address the contention problem in the entire memory space, we expose architecture features (i.e. physical address mapping for cache sets and DRAM banks) to OS by utilizing different types of addressing bits. Based on these architecture features, our ISCA-2014 paper introduces **Vertical Partitioning** (VP) [10] into modern OS. VP partitions the memory hierarchy vertically through cache and DRAM simultaneously to completely eliminate the contention issue for all susceptible levels in the memory hierarchy. (2) Combined with horizontal partitioning mechanisms (i.e, bank- or cache-only partitioning), we group the memory partitioning policies into several categories to form a memory management policy space for diverse workloads to choose from. This enables flexible and customized resource allocation that meets each individual application's memory and cache resource requirement. (3) By dynamically monitoring applications' page-table using a low-overhead algorithm, we equip the OS with the capability of "understanding" applications' memory behavior on the fly. Combining all these components and a large set of experimental results conducted in real systems, we devise an "intelligent" memory management mechanism that can choose appropriate allocation policy for workloads with arbitrary characteristics.

## 2. Solution

### 2.1 Vertical and Horizontal Partitioning

In memory controllers (MC), address mapping policies are determined by platform and configurations. Typically, beside the bits that only index DRAM banks (B-bits) and LLC sets (C-bits), there are also some bits that denote both (noted as O-bits). Figure 1 illustrates the three categories of coloring bits on a mainstream machine (Intel i7-860 with 8GB memory and 64 banks, B-bits: 21~22; C-bits: 16~18; O-bits: 14~15 [8,9,10]).

We observe that by considering different number of bits of various types when allocating a memory page, OS can exploit diverse memory allocation approaches. As shown in Table 1, bank-only partitioning and interleaving based page allocation can be derived by considering B-bits when allocating a page number to an application. Particularly, the O-bits enable vertical partitioning that partitions both LLC sets and DRAM banks vertically through the memory hierarchy. We further derive several sub-policies (i.e., A/B/C-VP) by choosing different combinations of O-, B- and C- bits to partition memory/cache into different quotas. Each policy represents one resource usage characteristic and has its own "friendly" (suitable) workloads, which performs better on this policy than on any other policies.

---
* I wish to extend my deep thanks to Yong Li (Pitt & VMware CA), Prof. Chen Ding (Rochester) and Prof. Xiaodong Zhang (Ohio) for their efforts.

*The first step work is published in ISCA-14 (Corresponding author: Lei Liu). Title: Going Vertical in Memory Management: Handling Multiplicity by Multi-Policy.*

| Policy | Coloring Bits | Description | Target Cores |
|---|---|---|---|
| Interleaving | B-bits {21~22} O-bits {14~15} | Bank-Interleaving w/ random page allocation | 4/8-core |
| Bank-Only | B-bits {21~22} O-bits {15} | LLC → 2 groups Banks → 8 groups | 4/8-core |
| A-VP | O-bits {14~15} | LLC → 4 groups Banks → 4 groups | 4-core |
| B-VP | B-bits {22} + O-bits {14~15} | LLC → 4 groups Banks → 8 groups | 8-core |
| C-VP | C-bits {16} + O-bits {14~15} | LLC → 8 groups Banks → 4 groups | 8-core |

**Table 1. Several representative partitioning policies**

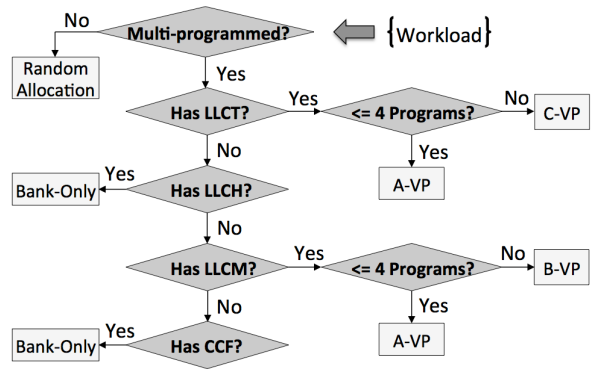

**Figure 2. Memory allocation policy decision tree (PDT)**

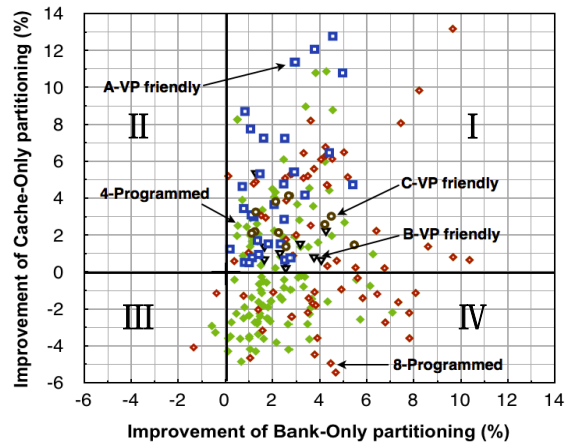

**Figure 3. Performance improvement of various polices for 214 workloads (A/B/C-VP are memory policies introduced in Table 1).**

### 2.2 Understanding Memory Features
#### 2.1.1 Key Observations and Classification
Based on numerous experiments across over 200 workloads in real system (Figure 3), we find that compared to DRAM, the amount of available cache resource has a much greater impact on application performance. Thus, we classify applications into four categories: Core Cache Fitting (CCF), LLC High (LLCH), LLC Middle (LLCM) and LLC Thrashing (LLCT), based on their performance drop caused by cache quota reduction from 8/8 (entire cache) to 1/8. CCF applications (e.g., *hmmer*), do not degrade significantly when using fewer LLC resources as their working sets fit into the L1 and L2 private caches. LLCT applications, such as *libquantum*, are also insensitive to cache quotas, but due to cache thrashing behavior rather than small working set sizes. LLCH applications such as *mcf* suffer the worst performance degradations from reduced cache quotas due to their large resource requirements.

#### 2.1.2 Dynamic Application Classification
To enable online optimization, we need to classify an application based on its run-time characteristics. **We found the number of *hot pages* (active pages used in a particular time interval and can be identified by the access bit in PTE) can reflect an application's LLC demand due to the DRAM row-buffer locality.** Based on this insight, we devise a classification algorithm implemented as two kernel tasks, JOB1 and JOB2, which sample the page access patterns periodically. JOB1 is responsible for collecting the number of hot pages by clearing the *access_bit* and examining pages with *access_bit* =1 at the end of each sampling period. JOB2 uses an array of page access counters to record the number of accesses for each page. JOB2 groups the counter values into ranges and computes a *weighted page distribution* (WPD) to reflect page reference locality. Our algorithm can accurately classify applications into one of the four categories detailed in Section 2.1.1 by comparing the number of hot pages and WPD with trained thresholds.

### 2.3 Memory Allocation Policy Selection
We adopt a data mining approach to quantitatively study the impacts of various memory allocation schemes on over 2000 workloads. We run each workload with different policies and record the performance improvements achieved by the corresponding policies. Based on the correlation between the classification and performance gains on different policies, we create a set of rules to select the policy of vertical management.

#### 2.3.1 Partitioning Rules
First, most workloads that contain at least one LLCT application perform best on A/C-VP. Second, a dominating percentage of workloads containing LLCH but not LLCT perform best on bank-only partitioning. Third, most workloads with LLCM but no LLCT or LLCH applications achieve best performance results with a modest cache partitioning scheme such as A-VP and B-VP. Based on the rules and their priorities relative to each other, we generate a memory management **policy decision tree (PDT)** shown in Figure 2. The PDT is useful for choosing appropriate policies for diverse workloads. Moreover, for multi-threaded workloads, Park et al. [12] argues that a random page-interleaved allocation scheme (Table 1) outperforms other schemes. Thus, we add this policy to handle multi-threaded workloads in PDT.

#### 2.3.2 Coalescing Rules
Despite the advantage in eliminating interference, a pure partitioning based approach is not always preferable since it limits the available resource and can harm the performance for resource hungry applications (e.g., LLCH). Thus, we extend the PDT with several coalescing rules that can be used to merge the partitioned resource quotas among certain types of applications. Using the data mining approach, we find the LLCH and LLCM applications should be coalesced together to share the cache quota, while LLCT and CCF applications should be coalesced respectively to share a small cache quota.

## 3. Experimental Results
We implement PDT with partitioning and coalescing rules in Linux kernel. Extensive experiments show that our framework outperforms the unmodified Linux kernel and achieves up to 11% performance gains over prior techniques. Our results also show that higher performance can be achieved by adaptively choosing workload's best-fit memory policy. Illustrated in Figure 3, half of the workloads fall into region 4, meaning their performance degrade in shared memory conditions where cache resource is being limited/partitioned, but bank partitioning can benefit them. By contrast, some workloads in region 1 can achieve above 10% performance gains via x-VP to eliminate interference.

## 4. Future Impact
DRAM and cache technology has been undergoing remarkable changes. In contrast to the fast-paced changes in the memory hierarchy, the legacy memory management strategies such as the order-based, interleaved memory/cache allocation schemes adopted in commodity OS and hardware remain largely unchanged. These existing strategies manage memory resource "blindly" in that they are not aware of the architecture features and applications' memory characteristics, leading to a more generic but less efficient approach. We develop a practical, cost-effective way to make the OS conscious about running applications and the underlying architecture, enabling a more adaptive and efficient way of utilizing memory resources. Our contributions may have the following long-term impacts on both academia and industry:

## 4.1 Academic Impact

**(1) New insight in memory optimization:** We conduct a comprehensive study in page-coloring based partitioning. We further propose a new vertical memory management mechanism to control the entire memory hierarchy, and thus eliminate the memory interference in the entire memory hierarchy. Our key insight is the overlapped address bits (O-bits) in physical address mapping. Using O-bits, we enable memory and cache vertical management that significantly mitigates the memory interference issue. Additionally, our idea to utilize architecture features (O-, B-, C- address mapping bits) enriches memory allocation policies and enlarges the space of memory optimization. Our study brings new opportunities and design patterns in the areas of high performance computing and memory architecture.

**(2) New method for application-aware computing/allocation:** Numerous prior studies demonstrate that many important system optimizations cannot be achieved without leveraging application behavior. In our work, we devise an online application classification method and implement it as Linux kernel modules. To the best of our knowledge, this is the first approach that captures dynamic application memory and cache usage patterns in real production settings, without the help of hardware based performance counters, long-running simulations, or pin-based profiling. The accuracy of our approach is verified by off-line profiling. Our work opens a new path for researchers to leverage the knowledge of running workloads for system optimizations.

**(3) New perspective for designing future OS:** To assist the application-aware policy selection process, we studied a large amount of workloads running on different memory allocation policies. Using data mining to analyze the results, we generate partitioning and coalescing rules used to appropriately partition resources while allowing non-interfering programs to live together for resource sharing. Our large result set provides valuable reference for studying the impact of memory scheduling/allocation methods on diverse workloads. In the long term, we believe that our approach, including (1) and (2), will motivate researchers in the related fields to build more intelligent modern OS that can understand and learn from application behavior to and adapt its own behaviors to maximize resource utilization and performance.

**(4) New approach for hybrid memory management.** We restructured the conventional memory strategy and firstly introduced the Vertical approach. We use this approach for hybrid DRAM-NVM memory management and devise Memos [19,20] in Linux kernel. Our first step results show memos can benefit the hybrid memory performance and the NVM lifetime.

## 4.2 Industry Impact

The benefits of our proposed vertical memory management to industrial world are multifold. (1) It adds both cache and DRAM into the OS management pool, and thus potentially benefits the overall system performance by simultaneously reducing cache and DRAM contention, a critical problem faced by many cloud providers such as Amazon, Google and VMware. (2) It significantly enlarges the memory management policy space and brings greater flexibility for diverse application needs in commercial data center and production environments. Moreover, application memory access and usage patterns are captured using a practical, page-table sampling based approach that only adds a very low overhead. (3) It helps reduce the energy cost and access latency of emerging NVMs. In particular, memory-partitioning techniques is more effective in NVM cases where row buffer miss latency and energy is larger. (4) It segregates applications with high latency-sensitivity versus those with bandwidth-sensitivity accesses (e.g., stream-like application), thus ensuring better QoS and fairness. Particularly, our partitioning and coalescing techniques can be used together to handle dynamic workload changes in production environments, thus having a profound influence on efficient resource isolation, virtualization and consolidation, which are critical and have a significant impact on the trend of "moving to the cloud".

By restructuring the buddy system in Linux kernel, we implement the HVR framework as an all-in-one solution that combines **H**orizontal, **V**ertical Partitioning and **R**andom allocation [12]. We believe our prototype demonstrates the feasibility of a more intelligent memory management strategy in modern OS design for addressing the emerging challenges in future complicated computing environments. It would require a minimal effort to port our prototype to production settings to benefit diverse commercial workloads.